# BloSEn: Blog Search Engine Based on Post Concept Clustering

S. Shanmugapriyaa[1], K.S.Kuppusamy[2], G. Aghila[3]

[1]Department of Computer Science, School of Engineering and Technology, Pondicherry University, Pondicherry, India

## ABSTRACT

*This paper focuses on building a blog search engine which doesn't focus only on keyword search but includes extended search capabilities. It also incorporates the blog-post concept clustering which is based on the category extracted from the blog post semantic content analysis. The proposed approach is titled as "BloSen (Blog Search Engine)" It involves in extracting the posts from blogs and parsing them to extract the blog elements and store them as fields in a document format. Inverted index is being built on the fields of the documents. Search is induced on the index and requested query is processed based on the documents so far made from blog posts. It currently focuses on Blogger and Wordpress hosted blogs since both these hosting services are the most popular ones in the blogosphere. The proposed BloSen model is experimented with a prototype implementation and the results of the experiments with the user's relevance cumulative metric value of 95.44% confirms the efficiency of the proposed model.*

## KEYWORDS

*Blogs, Crawler, Document Parser, Apache Lucene, Inverted Index, Clustering.*

## 1. INTRODUCTION

The Weblogs (Blogs) facilitates the end-user to share one's individual thoughts on the World Wide Web, without the need for any technical expertise. A typical blog combines text, images and other media along with links to other blogs and web pages related to its topic. It is a discussion site or an informational site where there are entries called posts. It induces deep discussion on a topic where the author shares some information on a topic and the public express their views and opinions as comments. This kind of interactivity distinguishes blogs from other static websites [19]. The ever growing data in the web are already in repository on the blogs. This search engine tends to effectively retrieve information and also display them to the user in a categorized way called Clustering. When we search information in static websites, we may not arrive correctly on the particular paradigm [16]. A blog, which is itself categorized into some topics, will have description of the same word under a defined paradigm. If that paradigm is the user's choice, then the user can go on with the results. Searching a blog in a normal web search manner will not yield effective result. A search engine specially made for searching blogs in depth is required to make user's search faster and effective [18].

The objectives are:

- To provide search based on every blog element in the entire archives of the blog posts.

- To display search results in a clustered manner based on the semantic categories retrieved from the blog post archives.

The paper is organized as follows: Section 2 is about the related works on blog search engines. Section 3 is about the model of the proposed blog search engine. It describes the various techniques used in the design. Section 4 discusses the efficiency of the algorithms used in the implementation of the proposed model. It also presents the user satisfaction and technical features reports. Section 5 is about conclusion and the future enhancements of this work.

## 2. RELATED WORKS

This section discusses the related works which motivated towards the proposed blog search engine. Works based on blogs belong to the Information Search and Retrieval domain. The proposed model includes the following fields of study:

- Searching

- Clustering

There are many blog search engines available in the web. One such blog search engine called Blog Ranger as mentioned in [1] provides multiple interfaces, each targeted at different goals. The goals of the work are mentioned as: topic search, blogger search and reputation search. Multi-faceted search means to show results from various viewpoints. Faceted Search as discussed in [15] addresses the weaknesses of conventional search approaches and users find it more effective information-seeking support.

The purpose of the proposed blog search engine is to explore the capabilities and limitations of existing weblog search engines. The features of a range of current blog search engines are described in [3]. These are then discussed and illustrated with examples that illustrate the reliability and coverage limitations of blog searching. The results illustrate blog search evaluation methods and do not use a full-scale scientific experiment. Some of the existing blog search engines as discussed in [6] are: (i) Google Blog Search which provides phrase control options, (ii) Technorati which provides related tags support, (iii) Ice Rocket which provides trend tool and spy support.

The dynamic nature of the blogosphere hinders the manual information extraction from it, promoting the development of new automated approaches. In [4], they proposed a component-based framework to create blog crawlers based on software architecture. This framework provides useful services for the blog analysis, including preprocessing, indexing, content extraction, classification, and tag recommendation.

The analysis of a large blog search engine query log, exploring a number of angles such as query intent, query topics and user sessions as discussed in [5]. The results show that the primary targets of blog searchers are tracking references to named entities and locating blogs by theme.

Clustering is the process of organizing the items based on set features. A study has proposed framework of categorizing blog posts according to their sub-topics. In the framework mentioned in [2], the sub-topic of each blog post is identified by utilizing Wikipedia entries as a knowledge source and each Wikipedia entry title is considered as a sub-topic label.

Apart from the above mentioned works, there were also works in Search Quality Criteria as discussed in [13] and summarization of search results as discussed in [20]. This paper collects the features from these related works and tries to bring out a blog search engine that is more efficient than the existing ones in terms of technical features as well as end-users needs.

## 3. THE BLOSEN MODEL

This section explores the proposed model of the blog search engine. The overview of the functions of the blog search engine is depicted in Fig. 1. The components in building the blog search engine are as follows: Crawler, Document Parser, Indexer and Query Processor. It makes use of Apache Lucene, which provides full text indexing and searching capabilities.

The crawler surfs through the blogs and extracts the archive links thus gathering the blog post pages as documents. The document content is parsed and it is stored in the Lucene document where the index is built. A search API is provided where query processing takes place based on the users' search words. The search API interacts with the Query Processor to process various kinds of queries and when results are retrieved, they are clustered based on categories. A separate log is maintained to record the users' search list.

This proposed model doesn't limit to the normal keyword search but also utilizes full text searching capabilities with the help of Apache Lucene [7].

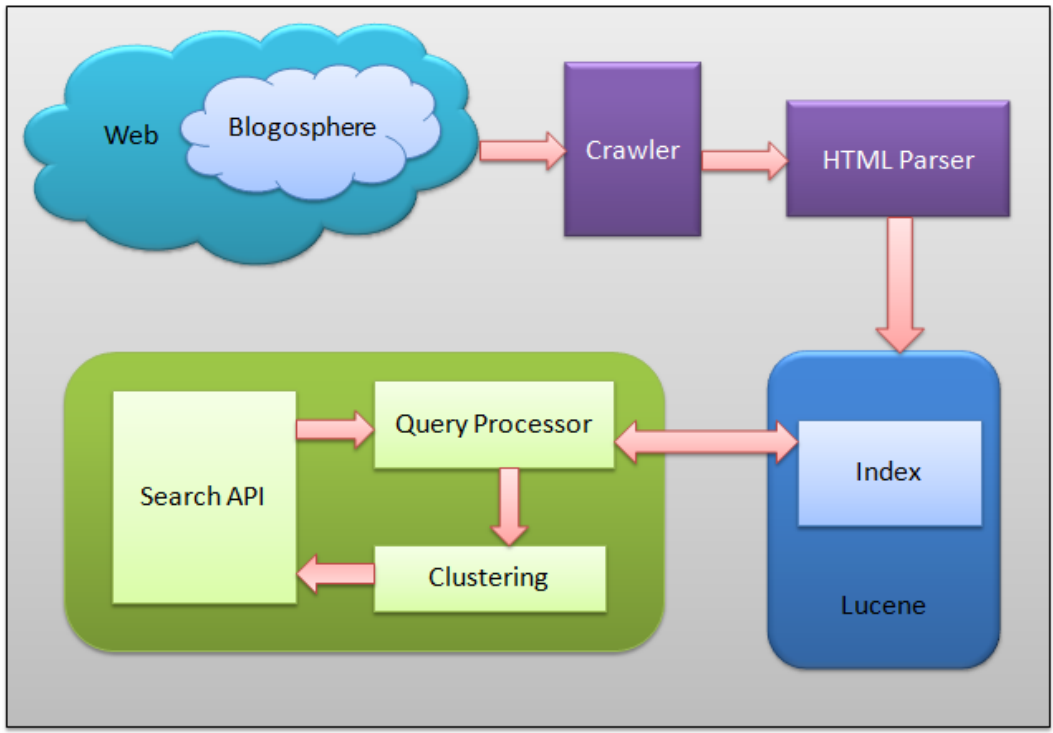

Figure 1. Block Diagram of proposed Model

When we get deep inside blog structure, there are many templates provided by different blog hosting services. Each template has its own defined structure which enables the programmers or developers to easily establish a particular design template for blogs in blog hosting services. Any kind of design that captures the user's attention will have the similar underlying template structure. Once we are familiar with this structure, we can easily differentiate the kinds of blogs.

International Journal of Ambient Systems and Applications (IJASA) Vol.1, No.3, September 2013

Once we differentiate the blogs, specific extraction technique can be applied to extract the blog elements without loss of data from the blogs. Search can be facilitated with the extracted elements and thus enabling users to find their requirements as soon as possible.

## 3.1 Components
### 3.1.1 Crawler
The crawler is the one which finds the links present in a web page. The crawler of this blog search engine finds archives links and individual post links. For this, it will undergo three levels of depth first search process. First level is to traverse the base URL, second level is to determine the archive URL and the final (third) level is to extract the individual post URL.

### 3.1.2 Parser
A document parser parses the documents and finds the important information. The post pages obtained by the crawler are converted as web page documents and then parsed. HTML parsing [9] is done to extract the blog elements such as blog url, blog title, blog name, blog generator service, post title, post url, postdate, post content, post author, post comments. Along with this, content analysis is done using Yahoo Content Analyzer [12] for the post content. All these elements are extracted for every individual post (page) of a blog. After extracting the elements, they are submitted to Apache Lucene for further process.

### 3.1.3 Apache Lucene

Apache Lucene [7] is an open source information retrieval software library that is widely recognized for its full text indexing and searching capabilities. The features of Lucene go as: Ranked searching, powerful query types, fielded searching, sorting by any field, multiple-index searching with merged results, allows simultaneous update and searching and configurable storage engine. The logical architecture of Lucene is a document containing fields of text. The fields can be both stored and indexed. The architecture is depicted in Fig. 2. It has majorly two functions. One is to store the fields in the document and the other is to index the same to facilitate search.

In Lucene, a Document is the unit of search and index. An index consists of one or more Documents. A Document consists of one or more Fields. A Field is simply a name-value pair. A field is a named sequence of terms. A term is a string. The same string in two different fields is considered a different term. Lucene allows us to perform queries on this index, returning results ranked by either the relevance to the query or sorted by an arbitrary field such as a document's last modified date. Lucene is able to achieve fast search responses because, instead of searching the text directly, it searches an index instead [12]. The extracted blog elements are stored in these fields and index is created over these fields. The index size is 20% to 30% of the text indexed. When a user searches a text, it is searched using the index as illustrated in [8] [10] and then those documents, which hold the text, are returned [11]. The index stores statistics about terms in order to make term-based search more efficient. Lucene's index falls into the family of indexes known as an inverted index. This is because it can list, for a term, the documents that contain it. This is the inverse of the natural relationship, in which documents list terms.

Lucene indexes may be composed of multiple sub-indexes, or segments. Each segment is a fully independent index, which could be searched separately. Indexes evolve by: creating new segments for newly added documents and merging existing segments. Lucene refers to documents by an integer document number.

Figure 2. Lucene Structure

### 3.2 The Algorithm

This section explores the algorithms in detail for the proposed model. There are 4 algorithms for the proposed blog search engine to work efficiently and the speciality of these algorithms is they are time-saving ones. Once they are time-saving and efficient, automatically the performance is also expected to be high.

The crawler algorithm used in this blog search engine saves time from crawling all the links present in a blog page.

```
Algorithm 1: Selective crawler

1. Visit the blog root url page
2. Fetch the archive links
3. Fetch archive link page
4. Extract post link from every archive page
```

The parser algorithm does a kind of semantic parsing using the structure of blogs based on blog hosting services.

```
Algorithm 2: HTML Parser

1. Extract the HTML content of the post page using
Jsoup library
2. Check whether template belongs to blogger or
wordpress
3. If blogger:
      3.1 Go with respective tags to extract elements
      from blog post page
      3.2 Extract post content and store temporarily
4. If wordpress:
      4.1 Go with respective tags to extract elements
      from blog post page
      4.2 Extract post content and store temporarily
5. Analyze the temporarily stored post content using
Yahoo content Analyzer API
6. Store the elements as fields in Apache Lucene and
index the same
```

The search algorithm is the crucial one which does searching and retrieval process.

```
Algorithm 3: Search

1. Enter the query in search box
2. Do full text search using the Lucene index created
by algorithm2
3. Get documents based on relevancy
4. Search a particular field or filter them by
archives if needed
```

The clustered display algorithm yields in the new way of displaying results other than traditional way.

```
Algorithm 4: Clustered Display

1. Fetch Lucene documents from search query
2. Get    the    categories    of    resulting
documents
3. Group the documents as per categories
4. Display them in tree-view structure
```

## 4. EXPERIMENTS AND RESULT ANALYSIS

The model is implemented with Java and JSP technologies in the Netbeans development tool using Apache Lucene library [7] and Jsoup HTML parsing library [9]. It also uses a API called Yahoo Content Analyzer [12] for analyzing the blog content. The resulting blog search engine produces results in a faster time and in an efficient way compared to the existing blog search engines [6]. It is because of the use of full text search capability that is given by Apache Lucene along with full text indexing capabilities. The results are displayed based on the ranking [14] created by relevance of the search text.

The results of the experiments are shown in Fig. 3 and Fig. 4. Fig. 3 corresponds to the clustered display of search results. The search interface is design user-friendly that supports not only keyword search but full text search [17]. Fig. 4 corresponds to the filter applied to the narrowed search in blogs. This results page is display in the traditional view as in other blog search engines.

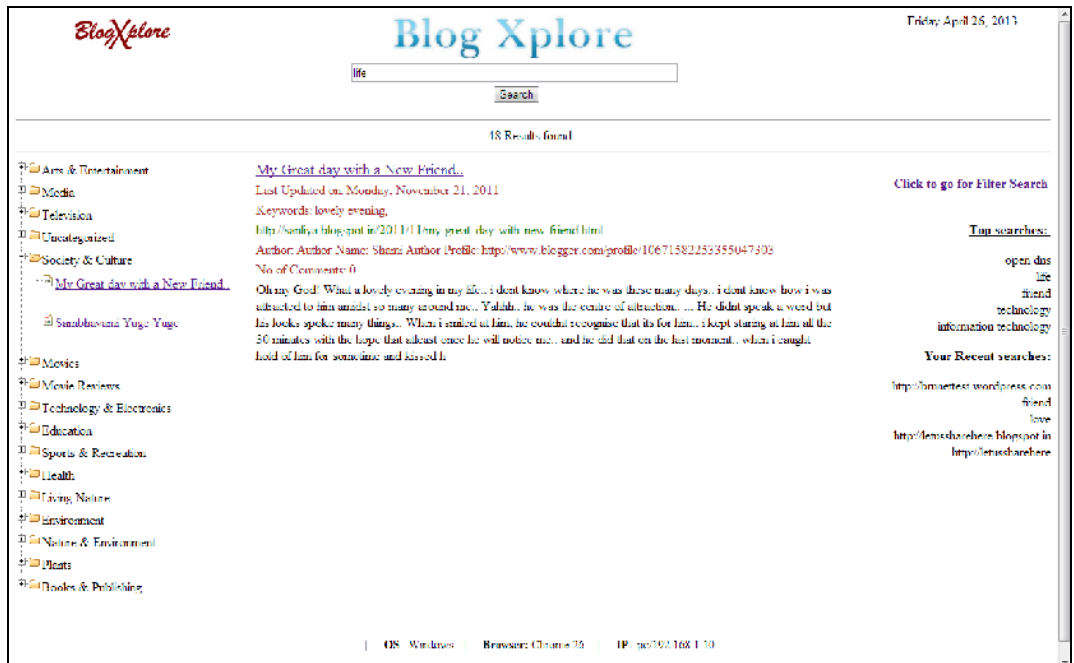

Figure 3. Clustered view display for results

International Journal of Ambient Systems and Applications (IJASA) Vol.1, No.3, September 2013

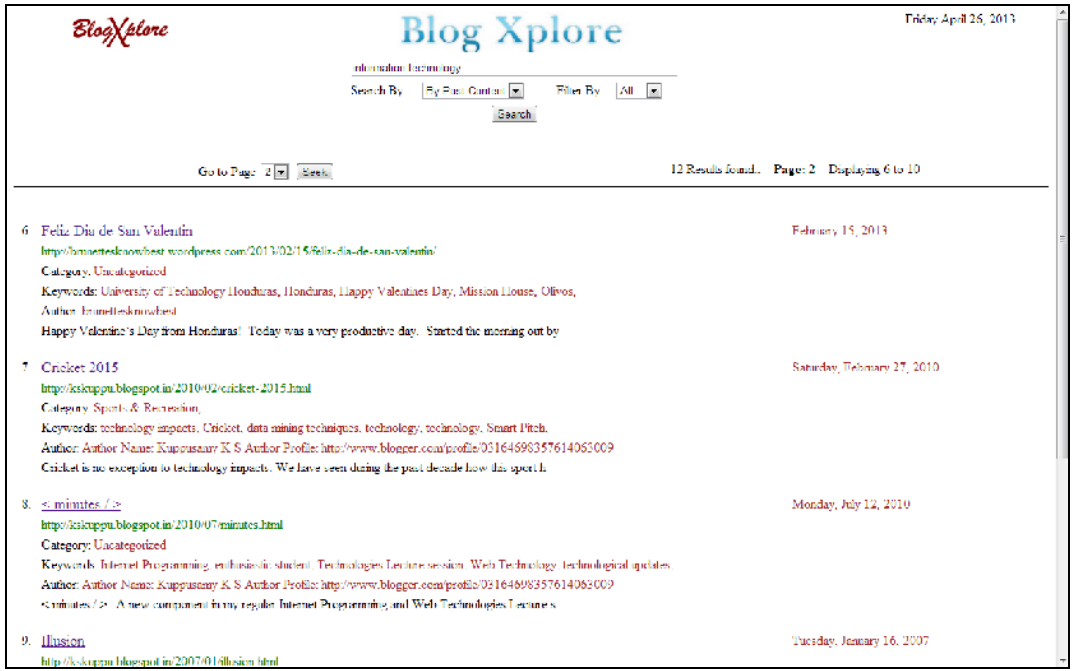

Figure 4. Traditional view display of results with filtered search

After the proposed model has been developed, they are tested for technical efficiency and for end-user satisfaction needs.

The technical features of the blog search engine are tabulated in Table 1. In every technical aspect, the implemented model proves to be efficient than the conventional ones.

Table 1: Technical features

| Technical Features | BloSen Implementation |
|---|---|
| Crawler | 3 levels of Depth First Search |
| Parser | Match Finding Technique |
| Content Analysis | Yahoo Content Analyzer API |
| Storage | Apache Lucene (Documents) |
| Index | Apache Lucene (Inverted Index) |
| Display of results | Clustered and Traditional view |

The end-user search features found in the experimented model that were absent in the existing blog search engine are summarized in the table below as follows:

Table 2: End-user features

| Features | Implemented model | Details | Existing search engines | Details |
|---|---|---|---|---|
| Normal Search | Yes | Search query | Yes | Search query |
| Traditional View | Yes | Link + snippet view + date + categories + keywords + author | Yes | Link + snippet view + date |
| Clustered View | Yes | Tree-view structure based on category with display of Link, snippet, title, date, author, keywords, comments | No | - |
| Navigation to all pages from every page | Yes | Drop down with all pages | No | Usual links with previous / next few pages |
| Search by Year Archives | Yes | Blogs are archived by years | No | - |
| Search by Date Range | No | Year archives work better than date range | Yes | Select from and to date range |
| Search by Filters for all fields | Yes | Phrase control / blog type / title / url / category / keyword / author | No | Only 3: Phrase control / author / title |
| Sort options | No | Sorts by relevance | Yes | Sort by date / relevance / popularity |
| User's Recent searches | Yes | Displays recent search list of a user | No | - |
| Top search list | Yes | Displays top searched queries | Yes | Some of the engines hold it |
| Details about user login | Yes | Displays OS, Browser and IP address of user's system | No | Hidden from user |

The BloSen prototype is experimented and the user's feedback is taken as a metric for evaluating the efficiency of the proposed model. The experiments were conducted in various sessions and the results of the experiments are tabulated in Table 3.

Table 3: BloSen User's Relevance Feedback

| Session ID | CRC | PRC | CIC |
|---|---|---|---|
| 1 | 90.5 | 4.7 | 4.8 |
| 2 | 93.5 | 4.2 | 2.3 |
| 3 | 87.3 | 5.6 | 7.1 |

| 4 | 87.1 | 8.2 | 4.7 |
| 5 | 81.5 | 10.1 | 8.4 |
| 6 | 80.1 | 16.5 | 3.4 |
| 7 | 83.5 | 9.6 | 6.9 |
| 8 | 89.1 | 6.5 | 4.4 |
| 9 | 91.3 | 6.6 | 2.1 |
| 10 | 93.5 | 4.7 | 1.8 |
| 11 | 92.1 | 3.2 | 4.7 |
| 12 | 93.1 | 4.5 | 2.4 |
| 13 | 92.1 | 3.5 | 4.4 |
| 14 | 92.4 | 3.4 | 4.2 |
| 15 | 84.4 | 8.8 | 6.8 |

In Table 3, CRC indicates Completely Relevant Category, PRC indicates Partially Relevant Category and CIC indicates completely irrelevant category. The data in Table 1 is illustrated in Figure 5 and Figure 6. It can be observed from the data that the mean of CRC across the sessions is observed as 88.7 % and the mean of PRC across the sessions is observed as 6.67%. The cumulative of CRC and PRC value of 95.44% confirms the efficiency of the proposed BloSen model.

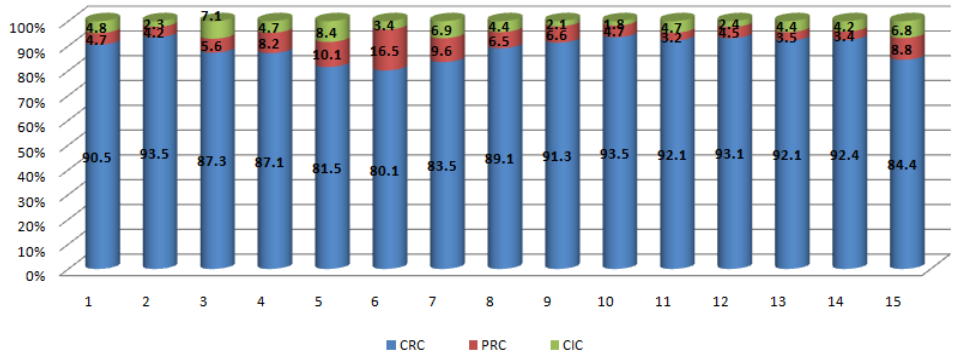

Figure 5. BloSen Relevance Comparison

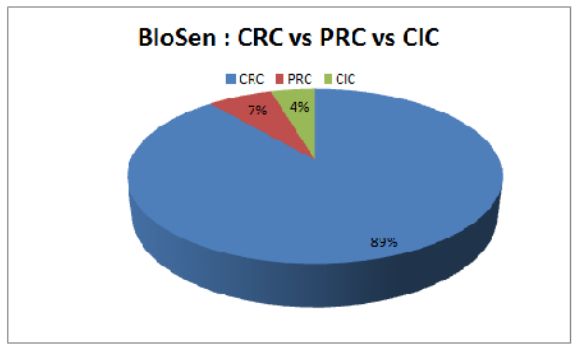

Figure 6. BloSen CRC vs PRC vs CIC

International Journal of Ambient Systems and Applications (IJASA) Vol.1, No.3, September 2013## 5. CONCLUSIONS AND FUTURE DIRECTIONS

Since there are many sources of information available to the users, the task of retrieving information becomes difficult when it comes to retrieving what the users exactly need. A Major chunk of the web is ruled by blogs in the present era. A search engine which typically explores blogs is very much essential. Thus, this implemented model enables the users to:

- The proposed BloSen model has achieved a cumulative user's relevance metric value of 95.44%.
- The BloSen model allows searching results based on various filtering criteria and fields.
- View clustered results and traditionally displayed results.
- Know his/her own recent searches and know the top searches done by other users.

Apart from this, the server also maintains the logs of every user who uses this search engine. Thus, this blog search engines combines many features of existing search engines and also provides full text searching capabilities with the use of latest technologies.

Future enhancements in this model that can be done are listed as:

- Extending this text search to multimedia search in blogs.
- Sub categories can be introduced to facilitate nested concept hierarchies instead of single level clustering.